\documentclass[11pt,a4paper]{article}
\usepackage{jheppub}
\usepackage{enumitem}
\usepackage{amsmath}
\usepackage{amsfonts}
\usepackage{graphicx}

\title{Near-Hagedorn Thermodynamics and Random Walks - Extensions and Examples}

\author[a]{Thomas G. Mertens,}
\author[a]{Henri Verschelde}
\author[b,c,d]{and Valentin I. Zakharov}
\affiliation[a]{Ghent University, Department of Physics and Astronomy\\
Krijgslaan, 281-S9, 9000 Gent, Belgium}
\affiliation[b]{ITEP, B. Cheremushkinskaya 25, Moscow, 117218 Russia,}
\affiliation[c]{Moscow Inst Phys \& Technol, Dolgoprudny, Moscow Region, 141700 Russia ,}
\affiliation[d]{School of Biomedicine, Far Eastern Federal University, Sukhanova str 8, 
Vladivostok 690950 Russia}

\emailAdd{thomas.mertens@ugent.be}
\emailAdd{henri.verschelde@ugent.be}
\emailAdd{vzakharov@itep.ru}
\abstract{In this paper, we discuss several explicit examples of the results obtained in \cite{theory}. We elaborate on the random walk picture in these spacetimes and how it is modified. Firstly we discuss the linear dilaton background. Then we analyze a previously studied toroidally compactified background where we determine the Hagedorn temperature and study the random walk picture. We continue with flat space orbifold models where we discuss boundary conditions for the thermal scalar. Finally, we study the general link between the quantum numbers in the fundamental domain and the strip and their role in thermodynamics.}

\begin{document}

\maketitle

\section{Introduction}
Despite its long history, high-temperature string theory remains a mysterious phase of nature that is not fully understood. The story in flat space goes roughly as follows. After heating a gas of closed strings, the constituents coalesce into (possibly several) long, highly excited closed string(s) \cite{Atick:1988si}\cite{Mitchell:1987hr}\cite{Mitchell:1987th}\cite{Bowick:1989us}\cite{Deo:1989bv}. The long string behaves as a random walker in a fixed spatial background \cite{Horowitz:1997jc}\cite{Barbon:2004dd}. From a Euclidean point of view, this long string is described by the thermal scalar. This is a complex scalar field that combines the winding $\pm 1$ stringy excitations around the Euclidean time circle. To what extent all of this generalizes to more general spacetimes is a question that was (largely) left unanswered in the past. \\

\noindent In a previous paper \cite{theory}, we analyzed the method set forth by the authors of \cite{Kruczenski:2005pj} to analyze the near-Hagedorn thermodynamics of string theory in general curved spacetimes directly from the string path integral. The method explicitly describes the random walk picture of high-temperature string thermodynamics. We noted in our previous paper that this random walk receives several corrections compared to the naive worldsheet dimensional reduction. In section \ref{pathderiv} we recapitulate these results from \cite{theory}. \\ 

\noindent In the main part of this paper, we will discuss several specific extensions and examples for which we will discuss the form of the thermal scalar action and the resulting near-Hagedorn random walk thermodynamics. These results illustrate the general discussion given in \cite{theory} and demonstrate the resolutions to some issues one might encounter in following the general treatment presented in section \ref{pathderiv}. \\
Firstly, in section \ref{dilatonbackground} we will discuss the linear dilaton background. This background is simple enough to be exactly solvable, yet it will teach us some valuable lessons about the interplay between dilatons in the spacetime action and their role in the string path integral. \\

\noindent In \cite{Grignani:2001ik}\cite{Grignani:2001hb}, the authors determine the Hagedorn temperature for a toroidally compactified flat spacetime geometry, when including constant metric and Kalb-Ramond background. This background has applications in non-commutative open string theory \cite{Gubser:2000mf}. The Hagedorn temperature in this background was determined by studying the divergence in the modular integral using the summation-of-particles strategy to evaluate the partition function. In section \ref{torcompmodel} we will apply our previous results to find the same critical temperature from the winding tachyon perspective. We also generalize this to the most general flat space toroidal model and determine the critical behavior.\\

\noindent Then, in section \ref{stringsinbox}, we consider orbifolds of flat space, where the orbifolding is done along one of the Cartesian coordinates. Such orbifolds are stringy models of strings-in-a-box. This generalization of the unorbifolded case will teach us an interesting fact regarding the link between string models and the associated thermal scalar particle models, in particular with respect to possible boundary conditions. \\

\noindent String thermodynamics in general can be rewritten in different ways: on the modular strip or on the fundamental domain. An interesting question is how the associated quantum numbers are related to corrections to thermodynamical quantities. For instance, higher winding modes and discrete momentum modes along the thermal circle on the fundamental domain do not represent physical (Lorentzian) states. Instead they represent corrections to thermodynamics. The precise sense in which this happens will be explored in section \ref{higherwind}, both for flat space and for a specific curved space example: the WZW $AdS_3$ model. \\

\noindent Several technical computations on toroidal models are given in the appendices.

\section{Set-up and plan}
\label{pathderiv}
The authors of \cite{Kruczenski:2005pj} have given an explicit path integral picture of the thermal scalar. We have extended their result in \cite{theory}. Let us first briefly review the arguments given there. We are interested in the torus path integral on the thermal manifold (obtained by Wick-rotating the time direction and periodically identifying with period $\beta$, the inverse temperature). After performing a modular transformation (as explained in \cite{theory}), we find the worldsheet action to be
\begin{eqnarray}
\label{action1}
S = \frac{1}{4\pi \alpha'} \left[ \left(1 + \frac{\tau_1^2}{\tau_2^2}\right) \int_0^{1/\tau_2} d\sigma \int_0^1 d\tau G_{\mu\nu} \partial_\sigma X^\mu \partial_\sigma X^\nu \right. \nonumber \\
\left.
+ 2 \frac{\tau_1}{\tau_2} \int_0^{1/\tau_2} d\sigma \int_0^1 d\tau G_{\mu\nu} \partial_\sigma X^\mu \partial_\tau X^\nu 
+ \int_0^{1/\tau_2} d\sigma \int_0^1 d\tau G_{\mu\nu} \partial_\tau X^\mu \partial_\tau X^\nu \right].
\end{eqnarray} 
To find the critical behavior, we focus on the $\tau_2 \to \infty$ contribution to the path integral and select the string state that is singly wound around the Euclidean time direction: $w = \pm 1$. We next perform a worldsheet Fourier series expansion:
\begin{align}
X^i(\sigma,\tau) & = \sum_{n=-\infty}^{+\infty} e^{i(2\pi n \tau_2) \sigma} X_n^i(\tau), \quad i=1\hdots D-1,\\
X^0(\sigma,\tau) & = \pm \beta \tau_2 \sigma +  \sum_{n=-\infty}^{+\infty} e^{i(2\pi n \tau_2) \sigma} X_n^0(\tau).
\end{align}
If we drop all higher Fourier modes and hence perform a naive worldsheet dimensional reduction, we found in \cite{theory} that the torus path integral on the thermal manifold reduces to
\begin{equation} 
Z_p = 2\int_0^\infty \frac{d \tau_2}{2\tau_2} \int \left[ \mathcal{D}X \right] \sqrt{\prod_{t} \det G_{ij}} \exp - S_p(X) 
\end{equation}
where 
\begin{equation}
\label{act}
S_p = \frac{1}{4\pi \alpha'}\left[ \beta^2 \int_0^{\tau_2} dt G_{00} + \int_0^{\tau_2} dt G_{ij} \partial_t X^i \partial_t X^j\right].
\end{equation}
The time parameter $t$ along the worldline is related to the worldsheet coordinate $\tau$ in (\ref{action1}) as $t=\tau_2 \tau$. \\
The full string partition function has been reduced to a partition function for a non-relativistic particle moving on the purely spatial submanifold. The time evolution of the particle in its random walk is identified with the spatial form of the long highly excited string. We view this as a realization of the Wick rotation: the long string in real spacetime has a form described by the above random walk.\\
The free energy of a gas of strings can then be identified with the single string partition function as \cite{Polchinski:1985zf}
\begin{equation}
F = -\frac{1}{\beta} Z_p.
\end{equation}
An alternative route we followed was the field theory of the thermal scalar. Using this theory, we were able to see that correction terms to the above particle action are in order. The thermal scalar action is given by (to lowest order in $\alpha'$)
\begin{equation}
\label{lowestFT}
S \sim \int d^{D-1}x \sqrt{G_{ij}}\sqrt{G_{00}}e^{-2\Phi}\left(G^{ij}\partial_{i}T\partial_{j}T^{*} + \frac{R^2G_{00}}{\alpha'^2}TT^{*} + m^2TT^{*}\right),
\end{equation}
where $m^2$ is the tachyon mass$^2$ in flat space whose precise value will be given below. From this action, the one-loop free energy is given by
\begin{align}
\label{FT}
\beta F &= - \int_{0}^{+\infty}\frac{dT}{T} \text{Tr} e^{-T\left(-\nabla^{2}+m_{\text{local}}^2 - G^{ij}\frac{\partial_{j}\sqrt{G_{00}}}{\sqrt{G_{00}}}\partial_{i}\right)} \\
\label{randwalk}
&= - \int_{0}^{+\infty}\frac{dT}{T}\int_{S^{1}} \left[\mathcal{D}x\right]\sqrt{G}e^{-\frac{1}{4\pi\alpha'}\int_{0}^{T}dt\left(G_{ij}(x)\dot{x}^{i}\dot{x}^{j} + 4\pi^2\alpha'^2\left(m_{\text{local}}^2 + K(x\right)\right)}.
\end{align}
We have also collected the `local' mass terms in
\begin{align}
\label{mlocal}
m_{\text{local}}^2 &= -\frac{4}{\alpha'} + \frac{R^2G_{00}}{\alpha'^2}, \quad \text{for bosonic strings}, \\
m_{\text{local}}^2 &= -\frac{2}{\alpha'} + \frac{R^2G_{00}}{\alpha'^2}, \quad \text{for type II superstrings}, \\
m_{\text{local}}^2 &= -\frac{3}{\alpha'} +\frac{1}{4R^2G_{00}}+\frac{R^2G_{00}}{\alpha'^2}, \quad \text{for heterotic strings}.
\end{align}
The function $K(x)$ denotes the following metric combination\footnote{$\nabla^2$ is the Laplacian on the spatial submanifold.}
\begin{equation}
\label{K}
K(x) =-\frac{3}{16}\frac{G^{ij}\partial_iG_{00}\partial_jG_{00}}{G_{00}^2} + \frac{\nabla^2 G_{00}}{4G_{00}}
\end{equation}
and this represents the effect of removing the $\sqrt{G_{00}}$ from the measure in the field theory action. Going from (\ref{FT}) to (\ref{randwalk}) requires some delicate manipulations that we discussed in \cite{theory}. Equation (\ref{randwalk}) can then be identified with (\ref{act}) and hence we can see which correction terms are needed in the particle action. \\
The correction terms are of three different types.
\begin{itemize}
\item{Firstly we have a correction term coming from the mass of the flat space tachyon and this is of the following form
\begin{equation}
\Delta S = -\frac{\beta_{H,\text{flat}}^2\tau_2}{4\pi\alpha'}.
\end{equation}
For bosonic strings $\beta_{H,\text{flat}}^2 = 16\pi^2\alpha'$, for type II superstrings $\beta_{H,\text{flat}}^2 = 8\pi^2\alpha'$ and for heterotic strings $\beta_{H,\text{flat}}^2 = 12\pi^2\alpha'$. 
For bosonic and type II strings, this is the flat space Hagedorn temperature but for heterotic strings this is not the case. By abuse of notation, we will nonetheless denote this term with $\beta_{H,\text{flat}}^2$.
}
\item{Secondly we have a correction coming from the $G_{00}$ component as explained in \cite{theory}: 
\begin{equation}
\label{corr}
\Delta S = \frac{1}{4\pi\alpha'}\int_{0}^{\tau_2}dt 4\pi^2\alpha'^2 K(x).
\end{equation}
}
\item{Finally we could have order-by-order $\alpha'$ correction terms of the field theory action (\ref{lowestFT}). These are of course not present in (\ref{randwalk}) and it is difficult to say anything specific about these for the general case. We have already analyzed their influence for two specific examples: Rindler space in \cite{Mertens:2013zya} and $AdS_3$ and $BTZ$ in \cite{Mertens:2014nca}.}
\end{itemize}
We also presented a simple extension to include a background NS-NS field. 
The string worldsheet action (\ref{action1}) has the following extra contribution
\begin{equation}
\label{KRexten}
S_{\text{extra}} = - i\frac{1}{2\pi\alpha'}\int_{0}^{1/\tau_2}d\sigma\int_{0}^{1}d\tau B_{\mu\nu}(X)\partial_{\sigma}X^{\mu}\partial_{\tau}X^{\nu}
\end{equation}
which results in the following augmentation of the particle action (\ref{act})
\begin{equation}
\label{KRexten2}
S_{\text{extra}} = \mp i\frac{\beta}{2\pi\alpha'}\int_{0}^{\tau_2}dt B_{0i}(X)\partial_{t}X^{i}.
\end{equation}
This action represents a minimal coupling of a point particle to a vector potential $A_{i} = B_{0i}$. This term breaks the symmetry between both windings because the NS-NS field breaks the orientation reversal symmetry of the string. From a point particle viewpoint, this means that the particles are oppositely charged under the electromagnetic field. \\

\noindent The generalization to stationary spacetimes was also given, resulting in
\begin{equation} 
\label{statio}
Z_p = 2\int_0^\infty \frac{d \tau_2}{2\tau_2} \int \left[ \mathcal{D} X \right] \sqrt{\prod_t \det \left(G_{ij} - \frac{G_{0i}G_{0j}}{G_{00}}\right)} \exp - S_p( X) 
\end{equation}
where 
\begin{equation}
S_p = \frac{1}{4\pi \alpha'}\left[ \beta^2 \int_0^{\tau_2} dt G_{00} +\int_0^{\tau_2} dt \left(G_{ij} - \frac{G_{0i}G_{0j}}{G_{00}}\right) \partial_t X^i \partial_t X^j\right].
\end{equation}
Like the static case described above, also this formula requires corrections, the simplest of which again being the correction for the flat space tachyon mass. \\

\noindent In \cite{theory}, we only analyzed flat Minkowski backgrounds explicitly, and hence we would like to see whether the above description really gives the correct answers in some explicit backgrounds. 
For a general curved background, we have much less control on what precisely happens. One of the problems we face is the exact description of the $\alpha'$ corrections to the thermal scalar action. Without these, we can not obtain a correct random walk description.\\

\noindent To avoid having to deal with this issue in full detail, we will, for the most part, focus on backgrounds that are geometrically trivial but with non-trivial topology. \\

\noindent In the next few sections, we will provide explicit solvable examples of this random walk picture. Each of these will highlight a different feature of the random walk (and the thermal scalar) description. 

\section{An exactly solvable model: linear dilaton background}
\label{dilatonbackground}
To start with, we will analyze the linear dilaton model. This will teach us some lessons on how to treat dilaton backgrounds in the thermal scalar formalism and also on the influence of continuous quantum numbers on the critical temperature. Besides being a grateful toy model, this background is relevant since (among others) the asymptotic region of the $SL(2,\mathbb{R})/U(1)$ black hole (and the near-extremal NS5 black hole) is equal to such a background (see e.g. \cite{Sugawara:2012ag}\cite{Giveon:2013ica} and references therein). We will see several consistency checks to our general earlier results \cite{theory}.

\subsection{Paradox concerning dilaton backgrounds}
There is seemingly a paradox when considering backgrounds with a non-trivial dilaton. In the gauge-fixed string path integral, the dilaton background is not present (because it multiplies the world-sheet Ricci-scalar), while in the field theory action it is always present (at least as a correction to the measure). This puzzle is of course well-known and while the dilaton does not appear explicitly in the string path integral, it does show up in worldsheet computations such as OPEs etc. It is interesting to see how this paradox is avoided in our case. We will restrict ourselves to the bosonic string. Consider flat space in $D<26$ dimensions ($G_{\mu\nu} = \delta_{\mu\nu}, B_{\mu\nu} = 0$). In order to have a consistent string background, we are required to include a linear dilaton, say in the $d$-direction:
\begin{equation}
\Phi = \sqrt{\frac{26-D}{6\alpha'}}X^{d}
\end{equation}
and this solution is exact in $\alpha'$. We analyze the near-Hagedorn one-loop thermodynamics from two points of view.\\

\noindent The dilaton does not appear directly in the particle path integral (\ref{act}). In \cite{theory} we computed explicitly the corrections to the flat space result that one obtains by integrating out (rather than setting to zero) the higher worldsheet Fourier modes. Since we are now considering $D<26$ dimensions, the exact correction to the particle path integral is given by
\begin{equation}
\label{lindil}
e^{4\pi\tau_2}e^{\frac{\pi\tau_2}{6}(D-26)}
\end{equation}
for the linear dilaton background, the first factor being the (universally present) correction from the flat space tachyon and the second factor is an additional contribution solely present when the number of (flat) dimensions is less than $26$.\\

\noindent Next we analyze this background from the field theory point of view. The field theory can be taken off-shell so let us consider all backgrounds turned off except a non-trivial dilaton (whose form is generic for now). A non-trivial dilaton appears in the measure of the field theory action and it can be treated in the same way as the $G_{00}$ metric component discussed above. Using the substitution $G_{00} \to e^{-4\Phi}$ in (\ref{K}), this gives an extra contribution to the resulting particle action given by
\begin{align}
K(x) &= -\frac{3}{16}\frac{(\nabla e^{-4\Phi})^2}{e^{-8\Phi}} + \frac{\nabla^2 e^{-4\Phi}}{4e^{-4\Phi}} \\
&= (\nabla \Phi)^2 - \nabla^2 \Phi.
\end{align}
Up to this point, we have not imposed the background to be a consistent string background. To match this to the string path integral result, we need to go on-shell (the dilaton should be a linear dilaton). This immediately yields
\begin{equation}
K(x) = -\frac{D-26}{6\alpha'}
\end{equation}
and the action gets the correction, according to (\ref{corr})
\begin{equation}
\Delta S = \frac{1}{4\pi\alpha'}\int_{0}^{\tau_2}dt 4\pi^2\alpha'^2K(x) = -\pi\tau_2\frac{D-26}{6}.
\end{equation}
The extra contribution to the associated path integral (\ref{FT}) is then finally given by
\begin{equation}
\exp\left(\pi\tau_2\frac{D-26}{6}\right)
\end{equation}
and this precisely matches the extra second factor in the path integral result (\ref{lindil}). \\

\noindent It is curious to realize that in the path integral the number of dimensions gives the correction, while in the field theory the dilaton itself gives the correction. These are only linked when going on-shell. As discussed above, this has been known a long time, but it is reassuring to see it arise in this context as well. We conclude that there is no contradiction between the string path integral and the field theory when including dilaton backgrounds as soon as we use the equations of motion of the background fields. This actually provides an explicit example of our statements in \cite{Mertens:2013zya}: the field theory point of view allows us to go off-shell and provides a natural off-shell generalization of the worldsheet results. 

\subsection{Critical behavior}
Despite the fact that the string genus expansion in this background is not a good approximation, we can still formally determine the Hagedorn temperature both from the partition function and from the thermal spectrum. We will see that these again match as we expect. This exercise will learn us something valuable concerning continuous quantum numbers in conformal weights. 
The Hagedorn temperature of the linear dilaton background can be readily computed (both from the string path integral and from the field theory point of view) and is given by
\begin{equation}
\label{haglindil}
\beta_{H}^2 = 4\pi^2\alpha'\left(\frac{D-2}{6}\right).
\end{equation}
As usual, this can also be seen in the thermal string spectrum. We define a tachyon as a state that causes the (one-loop) free energy to diverge. We describe this in a general way. In a general bosonic string CFT, the one-loop partition function is given by
\begin{equation}
Z = \int_{F}\frac{d\tau_1 d\tau_2}{2\tau_2} \text{Tr}\left[q^{L_0-c/24}\bar{q}^{\bar{L_0}-\bar{c}/24}\right] = \int_{F}\frac{d\tau_1 d\tau_2}{2\tau_2}\left|\eta(\tau)\right|^4(q\bar{q})^{-\frac{1}{12}}\sum_{H_{\text{matter}}}{q^{h_i-1}\bar{q}^{\bar{h_i}-1}}.
\end{equation}
In the second equality, we sum over only the matter contributions (of the full $c=26$ (or $c=10$) matter CFT). We have isolated a $q\bar{q}$ combination, since this precisely compensates the ghost CFT in its asymptotic behavior, meaning
 \begin{equation}
Z \to \int_{F}\frac{d\tau_1 d\tau_2}{2\tau_2}\sum_{H_{\text{matter}}}{q^{h_i-1}\bar{q}^{\bar{h_i}-1}}
\end{equation}
as $\tau_2 \to \infty$. 
A tachyonic state in bosonic string theory is thus determined if the conformal dimension $h+\bar{h}$ in the matter sector is smaller than 2 (divergence for $\tau_2 \to \infty$ in $Z$) after integrating over continuous quantum numbers.\footnote{By this we mean that when considering the expression
\begin{equation}
\sum_{H_{\text{matter}}}{q^{h_i-1}\bar{q}^{\bar{h_i}-1}},
\end{equation}
there can be continuous states in the matter Hilbert space. The integral over these continuous states needs to be performed \emph{before} concluding whether there is really a divergence. For type II superstrings, the conformal dimension $h+\bar{h}$ needs to be less than $1$ to have a tachyonic state.} These continuous quantum numbers can give a non-vanishing contribution if they integrate into a $\tau_2$-dependent exponential. This indeed happens for the linear dilaton background as we now illustrate. \\
A general string state in the thermal linear dilaton background has conformal weight
\begin{align}
h &= \frac{\alpha'}{4}\left(\frac{2\pi n}{\beta} + \frac{w\beta}{2\pi\alpha'}\right)^2 + \frac{\alpha'k_i^2}{4} + \frac{\alpha'k_{d}}{4}\left(k_{d}-2iV\right) + N, \\
\bar{h} &= \frac{\alpha'}{4}\left(\frac{2\pi n}{\beta} - \frac{w\beta}{2\pi\alpha'}\right)^2 + \frac{\alpha'k_i^2}{4} + \frac{\alpha'k_{d}}{4}\left(k_{d}-2iV\right) + \bar{N},
\end{align}
where $d$ denotes the linear dilaton direction, $i$ represents all other flat directions and $V = \sqrt{\frac{26-D}{6\alpha'}}$.
The thermal scalar ($w=\pm1$, $n=0$) then has conformal weight
\begin{equation}
h = \bar{h} = \frac{\beta^2}{16\pi^2\alpha'} + \frac{\alpha'k_i^2}{4} + \frac{\alpha'k_{d}}{4}\left(k_{d}-2iV\right).
\end{equation}
In the partition function, we integrate over the transverse $k_i$ (like in the flat space case). These do not influence the Hagedorn temperature since these integrate to a square root prefactor. In the $d$-direction however, the integration over continuous quantum numbers is important since the integration over $k_d$ effectively gives an extra contribution of $\frac{26-D}{24}$ to the conformal weight. Since in this case $h=\bar{h}$, setting the conformal weight equal to 1 gives
\begin{equation}
\frac{\beta_H^2}{16\pi^2\alpha'} + \frac{26-D}{24} = 1
\end{equation}
which is the same expression as (\ref{haglindil}). Note that the Hagedorn temperature becomes infinitely large when $D=2$, so no tachyonic instability can set in in this spacetime dimension. This was known already a long time for the non-thermal tachyon. Note also that we focused on spacelike linear dilatons ($D<26$). Lightlike and timelike linear dilatons would correspond to mixing between the thermal compactification and the linear dilaton direction. We do not want to study this since these linear dilaton backgrounds are not static and thus it is meaningless to study thermodynamics with these.\\

\noindent For completeness, we present the relevant formulae for the type II superstring, which can be obtained analogously. The linear dilaton is now given by
\begin{equation}
\Phi = \sqrt{\frac{10-D}{4\alpha'}}X^{d}.
\end{equation}
The worldsheet correction term obtained in the large $\tau_2$ limit equals
\begin{equation}
e^{2\pi\tau_2}e^{\frac{\pi\tau_2}{4}(D-10)},
\end{equation}
which leads to the Hagedorn temperature
\begin{equation}
\beta_H^2 = 4\pi^2\alpha'\left(\frac{D-2}{4}\right).
\end{equation}

\section{Toroidally compactified background model}
\label{torcompmodel}
In \cite{theory} we have discussed the result of the worldsheet dimensional reduction for topologically trivial spatial dimensions. We did not analyze what happens when spatial coordinates include topological identifications and it is this feature that we will analyze more explicitly in this section. To evade the discussions concerning the missed worldsheet contributions (as we summarized in section \ref{pathderiv}), we will take the background fields to be (almost) trivial in the following way. We consider flat Euclidean space where $X^1$ is a compact coordinate
\begin{equation}
X^1 \sim X^1 + 2\pi R_1.
\end{equation}
Moreover, we will introduce a constant (Euclidean) metric and Kalb-Ramond field of the following form \cite{Grignani:2001ik}\cite{Grignani:2001hb}
\begin{eqnarray}
\label{metriKR}
G_{\mu\nu}=\left[\begin{array}{cccc} 
1-A^2 & -iA & 0 & \hdots \\
-iA & 1 & 0 & \hdots \\
0 & 0 & 1 & \hdots \\
\hdots & \hdots & \hdots & \hdots  \end{array}\right], \quad\quad
B_{\mu\nu}=\left[\begin{array}{cccc} 
0 & -iB & 0 & \hdots \\
iB & 0 & 0 & \hdots \\
0 & 0 & 0 & \hdots \\
\hdots & \hdots & \hdots & \hdots  \end{array}\right].
\end{eqnarray}
where $A$ and $B$ are real constants. Note the factors of $i$ coming from the fact that we consider an Euclidean target space. Although utilizing the Euclidean metric for such spacetimes (with hence parts of the metric being imaginary) looks quite subtle, we will see that we nonetheless reproduce the correct results. This background will be a check on the Kalb-Ramond extension to the random walker in section \ref{pathderiv}. One cannot globally diagonalize the metric tensor without disrupting the periodicity of the coordinates (there is a topological obstruction, much like a constant Wilson loop around a compactified dimension). Such backgrounds can give quite non-trivial consequences for string thermodynamics (see e.g. \cite{Dienes:2012dc} for a recent analysis of the situation of heterotic and type I strings in backgrounds with constant Wilson loops around the thermal circle). As discussed in the introduction, such backgrounds are used in non-commutative open string theory \cite{Gubser:2000mf}\cite{Grignani:2001hb}. This background, despite being relatively simple to handle analytically, will demonstrate the usefulness of the path integral worldsheet dimensional reduction approach. \\
We will study all different types of closed string theories in this background near their Hagedorn temperature. Extra material on this model is provided in the appendices. In appendix \ref{spec} we search for a winding tachyon in the exact string spectrum. In \ref{sop} we check the correspondence between the free energy and the path integral on the thermal manifold. \\
Let us now follow the procedure outlined in section \ref{pathderiv} to obtain the critical behavior purely from the string path integral.

\subsection{Dominant Hagedorn behavior}
We will derive the dominant Hagedorn behavior from the path integral. In section \ref{pathderiv} we did not have a worldsheet instanton contribution for the $X^{i}$. Here we must include this for the $i=1$ component. The Fourier expansion we use is 
\begin{align}
\label{expan}
X^0(\sigma,\tau) & =   \pm \beta \tau_2 \sigma +  \sum_{n=-\infty}^\infty e^{i(2\pi n \tau_2) \sigma} X_n^0(\tau), \nonumber\\
X^1(\sigma,\tau) & =  w\beta_1\tau_2\sigma + n\beta_1\tau + \sum_{n=-\infty}^\infty e^{i(2\pi n \tau_2) \sigma} X_n^1(\tau), \nonumber \\
X^i(\sigma,\tau) & =  \sum_{n=-\infty}^\infty e^{i(2\pi n \tau_2) \sigma} X_n^i(\tau),\quad i=2\hdots D-1.
\end{align}
where $w$ and $n$ are new quantum numbers labeling the winding around the $X^1$ direction. We alert the reader that these quantum numbers have nothing to do with the thermal winding and momentum quantum numbers. This is a slightly more general setting than the one studied in section \ref{pathderiv}. For $X^{0}$ we again only take the winding $\pm 1$ contribution, since we expect this mode to dominate (and we have explicitly checked that indeed this mode becomes massless at the Hagedorn temperature in appendix \ref{spec}). \\
We insert this expansion into the action (\ref{action1}) supplemented by the Kalb-Ramond extension (\ref{KRexten}).
From the worldsheet instanton contributions, we obtain the following non-oscillator contributions (coming from the terms in front of the series present in equation (\ref{expan}))
\begin{eqnarray}
S_{\text{non-osc}} = \frac{1}{4\pi\alpha'}\left[\left(1+\frac{\tau_1^2}{\tau_2^2}\right)\tau_2\beta^2(1-A^2) + \left(1+\frac{\tau_1^2}{\tau_2^2}\right)\left(\mp 2iA\beta\beta_1\tau_2w+w^2\beta_1^2\tau_2\right) \right. \nonumber \\
\left. +2\frac{\tau_1}{\tau_2}\left(\mp i A\beta\beta_1 n + w \beta_1^2 n \right)+\frac{\beta_1^2n^2}{\tau_2}\right]\mp \frac{\beta\beta_1 nB}{2\pi\alpha'}.
\end{eqnarray}
All that remains is to take the large $\tau_2$ limit of the worldsheet instanton sum
\begin{equation}
\sum_{n,w}\exp\left(-S_{\text{non-osc}}\right).
\end{equation}
Taking $\tau_2 \to \infty$ gives immediately that the $w=0$ contribution dominates. For the summation over $n$ the opposite happens: all terms contribute equally. We can get a handle on this by doing a Poisson resummation in $n$ 
\begin{eqnarray}
\sum_{n}\exp\left(-\frac{\beta_1^2n^2}{4\pi\alpha'\tau_2} \pm \frac{2i\tau_1A\beta\beta_1n}{4\pi\alpha'\tau_2} \pm \frac{B\beta\beta_1n}{2\pi\alpha'}\right) \nonumber \\
=\left(\frac{\alpha'\tau_2}{R_1^2}\right)^{1/2}\sum_{m}\exp\left(-\frac{4\pi^3\alpha'\tau_2 m^2}{\beta_1^2} \pm \frac{2\pi A\tau_1\beta m}{\beta_1} - \frac{\tau_1^2A^2\beta^2}{4\pi\alpha'\tau_2} \right. \nonumber \\
\left. \mp \frac{2\pi i B\tau_2m}{\beta_1} + \frac{\tau_2B^2\beta^2}{4\pi\alpha'} + i\frac{\tau_1AB\beta^2}{2\pi\alpha'}\right).
\end{eqnarray}
The third, fifth and sixth term in the exponential are independent of $m$ and the fourth term is an imaginary contribution. The first term dominates when $m=0$. When $m \neq 0$, the first term pushes the exponential to zero, regardless of what the second term does. So the sum is dominated by $m=0$.\footnote{These arguments can be made mathematically more precise: one first Poisson resums the series in $n$ (before setting $w=0$). The resulting double series in $w$ and $m$ is readily shown to converge uniformly in both parameters. This allows the large $\tau_2$ limit to be taken in the summand and the conclusion is the same.}
In all, we get the following particle action
\begin{eqnarray} 
S_{\text{part}} =  \frac{1}{4\pi \alpha'} \left[\frac{\tau_1^2}{\tau_2}\beta^2 + \tau_2\beta^2(1-A^2) - \tau_2B^2\beta^2 - 2i\tau_1AB\beta^2 - \beta_{H,\text{flat}}^2\tau_2  \right. \nonumber \\
\left. \pm 2 \frac{\tau_1}{\tau_2} \beta \int_0^1 d\tau G_{00} \partial_\tau X^0 \pm 2 \frac{\tau_1}{\tau_2} \beta \int_0^1 d\tau G_{0i} \partial_\tau X^i +
\frac{1}{\tau_2}\int_0^1 d\tau G_{\mu\nu} \partial_\tau X^\mu \partial_\tau X^\nu \right].
\end{eqnarray}
The first line contains all non-oscillator contributions. The last term in the first line is the result of the exact integration of the higher oscillator modes. This is the same as before since after extracting the non-oscillator contributions determined above (encoding the topological non-trivial aspects of the space), the remainder is the same as flat space. The last line contains the zero-modes of the coordinate fields. \\
The metric is constant (and the target fields are periodic), so out of the three terms in the second line, only the final term contributes.
When integrating this term, one is free to do a coordinate redefinition and change it to globally flat space. This allows one to integrate out the $X^{0}$ field, and all that remains is the $X^{i}$ fields that are non-interacting and living in flat space.\\
The $X^0$ integration produces a factor
\begin{equation}
\frac{\beta}{\sqrt{4\pi^2\alpha'\tau_2}}.
\end{equation}
Finally we integrate over $\tau_1$ using
\begin{equation}
\label{tau1}
\int_{-\infty}^{+\infty}d\tau_1\exp\left(-\frac{\beta^2}{4\pi\alpha'\tau_2}\tau_1^2 + i\frac{AB\beta^2}{2\pi\alpha'}\tau_1\right) = \sqrt{\frac{4\pi^2\alpha'\tau_2}{\beta^2}}\exp\left(-\frac{A^2B^2\beta^2}{4\pi\alpha'}\tau_2\right).
\end{equation}
Putting everything together, we arrive at
\begin{equation} 
\label{resComp}
Z_p = 2 \int_0^\infty \frac{d \tau_2}{2\tau_2} \left(\frac{\alpha'\tau_2}{R_1^2}\right)^{1/2} \int \left[ \mathcal{D}X \right]\exp - S_p(X) 
\end{equation}
where 
\begin{equation}
S_p = \frac{1}{4\pi \alpha'}\left[ - 16\pi^2\alpha' \tau_2  + \tau_2\beta^2(1-A^2-B^2+A^2B^2)
+\int_0^{\tau_2} dt \partial_t X^i \partial_t X^i\right].
\end{equation}
Convergence is achieved if
\begin{equation}
\beta^2(1-A^2)(1-B^2) > 16\pi^2\alpha',
\end{equation}
so the Hagedorn temperature is given by
\begin{equation}
\label{Hagtemp}
T_{H} = T_{H,\text{flat}}\sqrt{(1-A^2)(1-B^2)},
\end{equation}
exactly as predicted by either the string spectrum \ref{spec} or the divergence in the free energy \cite{Grignani:2001ik}. \\
We now make a few comments concerning this result. 
\begin{itemize}
\item{We get three correction terms to the Hagedorn behavior: schematically denoted as $A^2$, $B^2$ and $A^2B^2$ terms. It is curious to note that all of these have a different origin.
The $A^2$ is present simply from the $G_{00}$ component in the string worldsheet action.
The $B^2$ term only appears after a Poisson resummation in the variable $n$.
Finally the $A^2B^2$ term appears in the $\tau_1$ integration in the end. This is the second time we see that the $\tau_1$ integration can give us a crucial correction to the critical behavior: for the heterotic string in flat space, we saw that this was also the case in \cite{theory}. Of course, from the perspective of second quantized thermal scalar field theory, there is no analog of $\tau_1$ and we expect that the thermal scalar action fully reproduces these results. We will not discuss the results from this point of view in this section.}
\item{Note that in expression (\ref{resComp}) there is a prefactor $\left(\frac{\alpha'\tau_2}{R_1^2}\right)^{1/2}$ present. What is the significance of this factor? \\
One easily sees that it precisely cancels the path integral contribution from the compact direction $X^1$. The interpretation is that given by \cite{Barbon:2004dd}.\footnote{Equations (3.6) and (3.7) in their paper.} The random walk now has one compact dimension and this changes the large $\tau_2$ form of a random walk precisely in the manner described by the previous prefactor. Note that this reinforces the interpretation of the Euclidean path integral as the random walk describing the long string.}
\item{In \cite{Grignani:2001ik} the critical behavior was analyzed from a partition function perspective in the strip. In this modular domain, a saddle point procedure is required to handle the $\tau_1$ integration. We have seen that here we reproduce their result in a technically easier way (i.e. without having to resort to saddle point methods). Also, when using the methods of \cite{Grignani:2001ik}, it appears a daunting task to extend the result to multiple toroidal dimensions (including constant backgrounds). We extend our method in appendix \ref{generalization} to the most general flat toroidal model (including arbitrary constant backgrounds $G_{mn}$ and $B_{mn}$) and we will find agreement with the thermal winding state found in the Euclidean spectrum. The Hagedorn temperature we find using the thermal scalar path integral formalism is:
\begin{equation}
T_{H} = T_{H,\text{flat}}\sqrt{G_{00} + G^{lk}B_{l0}B_{k0}}.
\end{equation}}
\item{For type II superstring theories, we expect the only difference to be that the Hagedorn correction is replaced by the superstring correction. So our prediction for the Hagedorn temperature is the same, except we use $T_{H,\text{flat}} = \frac{1}{\sqrt{8}\pi\sqrt{\alpha'}}$ now. This again agrees with the result in \cite{Grignani:2001ik}.}
\item{For the heterotic string theories, some aspects change. In \cite{theory}, we have observed that the Hagedorn correction for heterotic strings (containing the $\beta_{H,\text{flat}}^2$ contribution to the particle action) is instead the following
\begin{equation}
e^{\pi i \tau_1}e^{3\pi \tau_2}.
\end{equation}
This $\tau_1$ contribution modifies the $\tau_1$ integration (\ref{tau1}) into
\begin{eqnarray}
\int_{-\infty}^{+\infty}d\tau_1\exp\left(-\frac{\beta^2}{4\pi\alpha'\tau_2}\tau_1^2 + i\frac{AB\beta^2}{2\pi\alpha'}\tau_1 + i \pi \tau_1\right) \nonumber \\
= \sqrt{\frac{4\pi^2\alpha'\tau_2}{\beta^2}}\exp\left(-\frac{A^2B^2\beta^2}{4\pi\alpha'}\tau_2 - AB\pi\tau_2 - \frac{\pi^3\alpha'\tau_2}{\beta^2}\right).
\end{eqnarray}

Taking all contributions, we see that for convergence we need
\begin{equation}
\beta^2(1-A^2)(1-B^2) + 4\pi^2AB\alpha'+\frac{4\pi^2\alpha'^2}{\beta^2} \geq 12\pi^2\alpha'.
\end{equation}
One recognizes this immediately as the same equation we wrote down in \ref{spec} for the critical radius when a certain string state becomes massless. The solutions to the preceding equation are hence the same as the Hagedorn temperature we wrote down in \ref{spec}.
The Hagedorn temperature is thus equal to
\begin{equation}
\beta_{H}^2 = \frac{6 - 2AB + 2\sqrt{8-6AB+A^2+B^2}}{(1-A^2)(1-B^2)}\pi^2\alpha'.
\end{equation}
}
\end{itemize}

\subsection{Interpretation of divergences}
\label{interpre}
\begin{figure}[h]
\centering
\includegraphics[width=0.2\textwidth]{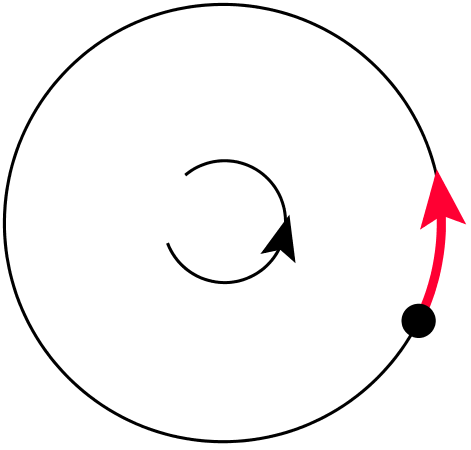}
\caption{For small $A^2$, timelike observers can remain stationary. For $A^2 \geq 1$, all observers must co-rotate with space (frame dragging effect), causing a breakdown of the field quantization along the Killing vector $\frac{\partial}{\partial t}$.}
\label{dragging}
\end{figure}

To finalize this section, let us give an interpretation of the critical values of $A$ and $B$ in this model. The toroidally compactified model has a peculiarity as $1-A^2=1$. We focus on only the $0$ and $1$ dimensions, for which the (Lorentzian signature) metric itself can be written in the standard stationary form:
\begin{equation}
ds^2 = -\alpha^2 dt^2 + (dx^1-Adt)(dx^1-Adt)
\end{equation}
with $\alpha = 1$. In the following we will use black hole terminology for convenience (despite there not being an actual black hole of course). \\
It is clear that there is no event horizon ($G_{00} - \frac{G_{01}G_{01}}{G_{11}} = -\alpha^2 \neq 0$), but there is indeed an ergoregion ($G_{00}= -(1-A^2) \stackrel{?}{=} 0$). This region is present if $A^2 \geq 1$ and is uniformly present all over space. One can interpret this as follows: space is a 1d interval with periodic identifications (a circle). With $A$ nonzero, absolute space is moving along this circle. When $A$ is sufficiently large, no timelike trajectory can remain stationary and the timelike Killing vector $\frac{\partial}{\partial t}$ that is used for quantization becomes spacelike everywhere. The situation is depicted in figure \ref{dragging}. 
In principle one can not discuss thermodynamics in terms of this observer anymore in this regime. This can also be seen in the results summarized in section \ref{pathderiv} (equation (\ref{statio})). For stationary spacetimes with Euclidean metric $G_{\mu\nu}$, one assumes both $G_{00} > 0$ and $G_{ij} - \frac{G_{0 i}G_{0 j}}{G_{00}}$ positive definite in the derivation. These conditions are related to the presence of ergoregions: $G_{00}$ vanishes at the stationary limit surface, whereas $\det\left(G_{ij} - \frac{G_{0 i}G_{0 j}}{G_{00}}\right)$ blows up there. \\ 

\noindent The $B=\pm1$ divergence on the other hand is related to the presence of a critical background field. One can relate the closed string gas in this $B$-field to an open string gas in a constant electric field around the compact dimension \cite{Grignani:2001ik}\cite{Grignani:2001hb}. For an electric field of too high strength, the string's tension cannot win from the electric field and the system becomes unstable towards string breaking. \\
We conclude by making an amusing remark. We have given intuitive explanations for both divergences. The $B$-divergence contains information on critical fields (related to the divergence of an open string gas in a constant electric field). The $A$-divergence on the other hand has to do with the spacelike behavior of the Killing vector used for describing thermodynamics. This divergence contains the information that the observer has to move faster than light when $A^2>1$ to remain stationary. It is interesting to again see that critical fields are T-dual to faster-than-light problems:\footnote{T-duality in the $X^1$-direction takes $A\leftrightarrow B$.} the DBI-action has a similar feature where the worldvolume electric fields have a critical value, T-dual to a D-brane moving faster than the speed of light. \\

\noindent We conclude that also in spaces with compact dimensions, we have a random walk behavior of the near-Hagedorn thermodynamical regime and we have derived this directly from the path integral. We have a prediction of the Hagedorn temperature in this space that agrees with both the string spectrum and with the divergence in the one-loop free energy.

\section{The thermal scalar on flat space orbifolds}
\label{stringsinbox}
A next question we ask is how boundaries affect the thermal scalar. In particular which boundary conditions should we impose at the boundary? To study this, we go one step further than the previous section and consider flat space orbifold compactifications. Because we are still geometrically in flat space, the string path integral is again exactly solvable. 

\subsection{The $S^1/\mathbb{Z}_2$ orbifold}
\begin{figure}[h]
\centering
\includegraphics[width=4cm]{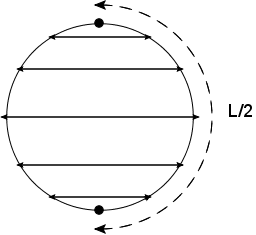}
\caption{The $S^1/\mathbb{Z}_2$ orbifold has two fixed points. It is obtained by a $\mathbb{Z}_2$ identification of a circular dimension.}
\label{orbif}
\end{figure}

Let us analyze the $S^1/\mathbb{Z}_2$ orbifold. We start with flat space (and all other background fields turned off) and make one dimension $X^1$ into a circle of circumference $L$. Then we impose the $\mathbb{Z}_2$ orbifold condition on this circle. This is a textbook model (see figure \ref{orbif}). 
First we will write down the near-Hagedorn limit of the exactly known partition function. \\
The situation outlined above corresponds to the following one-loop partition function \cite{Polchinski:1998rq}\cite{Burgess:1988qs}:
\begin{equation}
Z = Z(++) + Z(+-) + Z(-+) + Z(--),
\end{equation}
where
\begin{equation}
\label{untwisted}
Z(++) = \beta V\int_{F}\frac{d\tau_1 d\tau_2}{2\tau_2}\frac{1}{(4\pi^2\alpha'\tau_2)^{13}}\left|\eta(\tau)\right|^{-48}\sum_{p,q,n,m}\exp\left(- L^2\frac{\left|p-q\tau\right|^2}{4\pi\alpha'\tau_2} - \beta^2\frac{\left|m-n\tau\right|^2}{4\pi\alpha'\tau_2}\right)
\end{equation}
and
\begin{align}
\label{twisted}
&Z(+-) + Z(-+) + Z(--) =  \nonumber \\
&\beta A\int_{F}\frac{d\tau_1 d\tau_2}{2\tau_2}\frac{1}{(4\pi^2\alpha'\tau_2)^{25/2}}\left|\eta\right|^{-46}\left(\left|\frac{\eta}{\vartheta_2}\right| + \left|\frac{\eta}{\vartheta_3}\right| + \left|\frac{\eta}{\vartheta_4}\right| \right)\sum_{n,m}\exp\left(-\beta^2\frac{\left|m-n\tau\right|^2}{4\pi\alpha'\tau_2}\right).
\end{align}
The labels $+$ and $-$ correspond to path integral boundary conditions as we will discuss further on. In these formulas $V=A\frac{L}{2}$ where $A$ is the volume of all other (non-circular) dimensions. The quantum number $m$ and $n$ are the two wrapping numbers for the temporal dimension whereas $p$ and $q$ are the wrapping numbers for the spatial $X^1$ dimension. \\
After using the theorem of \cite{McClain:1986id}\cite{O'Brien:1987pn} to trade the fundamental domain for the strip (and dismissing the $n$-summation) and using the modular transformation $\tau \to -1/\tau$, the large $\tau_2$ limit becomes for the (+,+)-sector:
\begin{align}
\label{exact1}
&\iint_{\mathcal{A}} \frac{d\tau_1 d\tau_2}{\tau_2}\frac{\beta V}{(4\pi^2\alpha'\tau_2)^{13}}e^{4\pi\tau_2}\sqrt{\frac{4\pi^2 \alpha'\tau_2}{L^2}}e^{-\frac{\beta^2\left|\tau\right|^2}{4\pi\alpha'\tau_2}} = \iint_{\mathcal{A}} \frac{d\tau_1 d\tau_2}{2\tau_2}\frac{\beta A}{(4\pi^2\alpha'\tau_2)^{25/2}}e^{4\pi\tau_2}e^{-\frac{\beta^2\left|\tau\right|^2}{4\pi\alpha'\tau_2}},
\end{align}
and for the other sectors:
\begin{align}
\label{exact2}
\iint_{\mathcal{A}}\frac{d\tau_1 d\tau_2}{\tau_2}\frac{\beta A}{(4\pi^2\alpha'\tau_2)^{25/2}}&\left(\frac{e^{\pi\tau_2/6}}{2} + e^{-\pi\tau_2/12} + e^{-\pi\tau_2/12}\right)e^{23\pi\tau_2/6}e^{-\frac{\beta^2\left|\tau\right|^2}{4\pi\alpha'\tau_2}} \nonumber \\
&\to \iint_{\mathcal{A}}\frac{d\tau_1 d\tau_2}{2\tau_2}\frac{\beta A}{(4\pi^2\alpha'\tau_2)^{25/2}}e^{4\pi\tau_2}e^{-\frac{\beta^2\left|\tau\right|^2}{4\pi\alpha'\tau_2}},
\end{align}
upon dropping the $m=0$ temperature-independent contribution. By $\mathcal{A}$ we denoted the image of the modular transform $\tau \to -1/\tau$ on the strip region, which was shown in \cite{Kruczenski:2005pj}\cite{theory}. The important part is that the $\tau_1$ integral is effectively from $-\infty$ to $+\infty$ in the large $\tau_2$ region, and one can hence readily evaluate it. Note that a factor of 2 appears due to the sum over both windings.\\

\noindent Let us now analyze this background from the string path integral point of view along the lines of \cite{theory} as summarized in section \ref{pathderiv} and see whether we reproduce the above partition function. After the modular transformation $\tau \to -1/\tau$, the boundary conditions in the $X^1$ direction are
\begin{align}
X^{1}(\sigma+1/\tau_2,\tau) = \alpha X^{1}(\sigma,\tau) + p L, \\
X^{1}(\sigma,\tau+1) = \gamma X^{1}(\sigma,\tau) + q L,
\end{align}
where $\alpha$ and $\gamma$ are $\pm 1$ and represent the twisting along both torus cycles. The integers $p$ and $q$ represent the winding along both cycles. Not all of these sectors are present: in the twisted sectors with $\alpha = -1$ we have $p=q=0$. In the sector where $\alpha = 1$ and $\gamma = -1$, the $p$ and $q$ are also seen to vanish in the canonical formulation, although a priori this is not a requirement. So in all, we consider the following sectors: $(+,+,p,q)$, $(+,-,p,q)$, $(-,+,0,0)$ and $(-,-,0,0)$. We will comment on the $(+,-)$ sector further on. \\
To start, we do the worldsheet Fourier series expansion as in \cite{theory}. The $\alpha = 1$ sectors are the same as before. For the $\alpha = -1$ sectors on the other hand, this is the following expansion:
\begin{align}
X^0(\sigma,\tau) & = \pm \beta \tau_2 \sigma +  \sum_{n=-\infty}^{+\infty} e^{i(2\pi n \tau_2) \sigma} X_n^0(\tau),\\
X^1(\sigma,\tau) & = \sum_{n=-\infty+1/2}^{+\infty+1/2} e^{i(2\pi n \tau_2) \sigma} X_n^i(\tau), \\
X^i(\sigma,\tau) & = \sum_{n=-\infty}^{+\infty} e^{i(2\pi n \tau_2) \sigma} X_n^i(\tau), \quad i=2\hdots D-1.
\end{align}
In the $X^{1}$ direction, half-integer modes are used. These do not have a zero-mode and are subdominant in the $\tau_2 \to \infty$ limit. These modes are the twisted sector states that are localized in the $X^{1}$ dimension. We thus drop the $\alpha = -1$ sectors.\\
Next is the sum over toroidal windings $p$ and $q$. These are dealt with just like in the toroidal model discussed in section \ref{torcompmodel}: $p=0$ from the start. \\
After this step, we have only a sum over two sectors remaining: $\gamma = \pm 1$ (and the sum over $q$). This quantum number sets the boundary conditions on the point particle path integral in the $X^1$-dimension: $X^1(\tau_2) = \gamma X^1(0) + q L$. 
Explicitly we have for the $\gamma = 1$ sector:
\begin{align}
\label{untwist}
\sum_{q\in \mathbb{Z}} \int_{X^1(0) = X^1(\tau_2) + qL} \left[\mathcal{D}X^1\right]e^{-\frac{1}{4\pi\alpha'}\int_{0}^{\tau_2}dt\left(\partial_tX^1\right)^2} &= \int_{0}^{L/2}dx \sum_{q\in \mathbb{Z}} e^{-\frac{L^2q^2}{4\pi\alpha'\tau_2}}\frac{1}{\sqrt{4\pi^2\alpha'\tau_2}} \nonumber \\
&\to \quad \frac{1}{2},
\end{align}
and for large $\tau_2$ this yields simply $1/2$.
The integration over the zero-mode is only from $0$ to $L/2$ because the particle lives in a box of length $L/2$. The $\gamma = -1$ sector gives analogously:
\begin{equation}
\label{twist}
\sum_{q\in \mathbb{Z}} \int_{X^1(0) = -X^1(\tau_2) + qL} \left[\mathcal{D}X^1\right]e^{-\frac{1}{4\pi\alpha'}\int_{0}^{\tau_2}dt\left(\partial_tX^1\right)^2} = \int_{0}^{L/2}dx \sum_{q\in \mathbb{Z}} e^{-\frac{(x+qL/2)^2}{\pi\alpha'\tau_2}}\frac{1}{\sqrt{4\pi^2\alpha'\tau_2}},
\end{equation}
which becomes\footnote{Note that in \cite{Polchinski:1998rq} only the (+,+)-sector has non-zero $p$ or $q$. The last integral looks like a sector with only zero $q$, but where the final zero-mode integral runs over the entire volume. This explicitly shows that this term is independent of $L$ as in \cite{Polchinski:1998rq}.} 
\begin{equation}
\int_{0}^{L/2}dx \sum_q e^{-\frac{(x+qL/2)^2}{\pi\alpha'\tau_2}}\frac{1}{\sqrt{4\pi^2\alpha'\tau_2}} = \int_{-\infty}^{+\infty}dx e^{-\frac{x^2}{\pi\alpha'\tau_2}}\frac{1}{\sqrt{4\pi^2\alpha'\tau_2}} = \frac{1}{2}.
\end{equation}
Finally including also $X^0$ and the other uncompactified dimensions we find agreement with the previous expressions (\ref{exact1}) and (\ref{exact2}). When we take one step back, we have for the path integral in the $X^1$ direction:
\begin{equation}
Z_{X^1} = \int_{0}^{L/2}dx\left\{\sum_q e^{-\frac{L^2q^2}{4\pi\alpha'\tau_2}}\frac{1}{\sqrt{4\pi^2\alpha'\tau_2}} + \sum_q e^{-\frac{(2x+qL)^2}{4\pi\alpha'\tau_2}}\frac{1}{\sqrt{4\pi^2\alpha'\tau_2}}\right\}.
\end{equation}
This can be interpreted as a point particle in a box corresponding to Neumann boundary conditions for a scalar field at the boundary $X^1 = 0$ and $X^1 = L/2$ \cite{Kleinert:2004ev}.

\subsubsection*{Alternative view on the boundary conditions}
One can also see that Neumann boundary conditions are the only possibility directly from the point particle perspective. Consider a non-relativistic point particle of mass $m$ in a box of length $L/2$. We are interested in the (Euclidean) propagation amplitude for a scalar spin 0 particle starting at any point in the box and returning at the same point in the limit of large propagation time $\tau_2$. Consider then Dirichlet and Neumann boundary conditions for this point particle in the box. Dirichlet boundary conditions lead to 
\begin{align}
\int \left[\mathcal{D}X\right]&e^{-\frac{m}{2}\int_{0}^{\tau_2}dt\left(\partial_tX\right)^2} 
= \int_0^{\frac{L}{2}} dx\frac{4}{L}\sum_{n=1}^{+\infty}\sin\left(\frac{2\pi n x}{L}\right)^2e^{-\frac{4\pi^2n^2}{2mL^2}\tau_2} \quad \to \, e^{-\frac{\pi^2}{2mL^2}\tau_2},
\end{align}
whereas Neumann boundary conditions give
\begin{align}
\int \left[\mathcal{D}X\right]&e^{-\frac{m}{2}\int_{0}^{\tau_2}dt\left(\partial_tX\right)^2} 
= \int_0^{\frac{L}{2}} dx\sum_{n=0}^{+\infty}\left(\frac{4}{L}-\delta_{n,0}\frac{2}{L}\right)\cos\left(\frac{\pi n x}{L}\right)^2e^{-\frac{4\pi^2n^2}{2mL^2}\tau_2} \quad \to \, 1.
\end{align}
We know that changing spacetime solely in the spatial dimensions should not change the Hagedorn temperature, an example of which we have seen in section \ref{torcompmodel}. We notice that Dirichlet boundary conditions give a Hagedorn-like correction and hence change the Hagedorn temperature. This corresponds in the point particle model to the zero-point energy of a particle in a box. Neumann boundary conditions do not have this problem and it is clear from this point of view that indeed Neumann boundary conditions are the right choice.\\
Obviously, all of the results in this section readily extend to multiple boxed dimensions when making the required modifications.

\subsection{The $\mathbb{R}/\mathbb{Z}_2$ orbifold}
\begin{figure}[h]
\centering
\includegraphics[width=4cm]{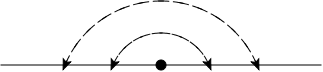}
\caption{The $\mathbb{R}/\mathbb{Z}_2$ orbifold obtained by taking $L \to \infty$ in the previous model.}
\label{orbif2}
\end{figure}

Taking $L \to \infty$, we arrive at the $\mathbb{R}/\mathbb{Z}_2$ orbifold (figure \ref{orbif2}).
Note that it is important to take $L\to\infty$ first and only then consider the large $\tau_2$ limit. For the $(+,+)$ sector we find using (\ref{untwist}):
\begin{equation}
\frac{L}{2} \frac{1}{\sqrt{4\pi^2\alpha'\tau_2}}
\end{equation}
where $L$ is the length before orbifolding (from $-\infty$ to $+\infty$) and the physical length remains $L/2$. The twisted sector computation (\ref{twist}) was exact in $\tau_2$ and remains the same (since it is independent of $L$).\footnote{We mention a technicality: taking $L \to \infty$ before rewriting the sum over $q$ as an integral, we retain the sectors $q=0$ and $q=-1$. The second term can be rewritten as an integral from $-L/2$ up to $0$ with the same integrand as the $q=0$ term. Then taking $L\to\infty$ yields the same result as before. One should be careful not to forget the $q=-1$ sector.} We notice that the $(+,+)$ sector is different than before and this has now a volume-dependence. This is consistent with the exact results, obtained by taking the large $L$ limit in (\ref{untwisted}) and (\ref{twisted}) and then performing the $\tau_2 \to \infty$ limit in the modular transformed domain:
\begin{align}
&\iint_{\mathcal{A}} \frac{d\tau_1 d\tau_2}{\tau_2}\frac{\beta V}{(4\pi^2\alpha'\tau_2)^{13}}e^{4\pi\tau_2}e^{-\frac{\beta^2\left|\tau\right|^2}{4\pi\alpha'\tau_2}}, \quad (+,+) \text{ sector}, \\
&\iint_{\mathcal{A}}\frac{d\tau_1 d\tau_2}{2\tau_2}\frac{\beta A}{(4\pi^2\alpha'\tau_2)^{25/2}}e^{4\pi\tau_2}e^{-\frac{\beta^2\left|\tau\right|^2}{4\pi\alpha'\tau_2}}, \quad \text{sum of other sectors}.
\end{align}

\subsection{Identification of the resulting particle models}
We have seen several different string approaches to making a compact model of strings in a box: toroidal models in section \ref{torcompmodel} and orbifold models in this section. We have seen that the orbifold model actually corresponds to a particle in a box with Neumann boundary conditions. Let us now make explicit the analogous particle model for the toroidally compactified string models. For the toroidal compactification discussed in the previous section (but this time \emph{without} background fields), we have the expansion
\begin{equation}
X^1(\sigma,\tau) \approx w \beta_1 \tau_2 \sigma + n \beta_1 \tau + X^{1}_{0}(\tau),
\end{equation}
where we again only retained the lowest Fourier mode which is periodic ($X^{1}_0(0) = X^{1}_0(1)$). Taking $\tau_2 \to \infty$ previously gave us that $w=0$. Instead of choosing the periodic coordinate $X^{1}_0$, we regroup the field into
\begin{equation}
X^1(\sigma,\tau) \approx X^{1}(\tau),
\end{equation}
where this last field has the property that $X^{1}(1) = nL + X^{1}(0)$ (with $L=\beta_1$). Although a trivial step, this means that there is no modification compared to the uncompactified case exhibited in section \ref{pathderiv}, except for this boundary condition. So in the end we get a particle path integral with these boundary conditions. One recognizes this immediately as a particle path integral on a circle \cite{Kleinert:2004ev} since the path integral is given by
\begin{equation}
\sum_{n=-\infty}^{+\infty}\int_{X^1(\tau_2) = X^1(0) + nL} \left[\mathcal{D}X^1\right]e^{-\frac{1}{4\pi\alpha'}\int_{0}^{\tau_2}dt\left(\partial_tX^1\right)^2} = \int_0^{L} dx\frac{1}{L}\sum_{n=-\infty}^{+\infty}\exp\left(-\frac{4\pi^3\alpha'n^2}{L^2}\tau_2\right).
\end{equation}
It asymptotes to $1$ when $\tau_2 \to \infty$ and this is the value we previously found (the Poisson dual variable was equal to zero).\footnote{See the second comment after equation (\ref{Hagtemp}).} Just like in the orbifold model, we see that as we take the $\tau_2 \to \infty$ limit in the worldsheet path integral, we can identify the intermediate expressions as point particle path integrals. In this point particle path integral, we need to take the same $\tau_2 \to \infty$ limit again, and this gives us finally the limiting Hagedorn behavior. Summarizing, we have the following correspondences between string models and particle models for the near-Hagedorn thermodynamics of the thermal scalar:
\begin{align}
\text{Toroidal model} \quad &\Leftrightarrow \quad \text{Point particle on circle}, \nonumber \\
\text{Strings in a box = } S^1/\mathbb{Z}_2 \quad &\Leftrightarrow \quad \text{Point particle in a box with Neumann BC}, \nonumber \\
\text{Strings in a halfspace = } \mathbb{R} / \mathbb{Z}_2 \quad &\Leftrightarrow \quad \text{Point particle in a halfspace with Neumann BC}. \nonumber
\end{align}
We conclude that string theory in its critical regime gives a particle theory of the thermal scalar whose boundary conditions are entirely dictated by the full string theory \cite{Burgess:1988qs}. One cannot choose other boundary conditions freely.

\section{Higher winding modes and multistring states}
\label{higherwind}
In this section we look at an entirely different question concerning string thermodynamics. It is known that the strip and fundamental domain both are viable routes to string thermodynamics. The main question we would like to examine here is how precisely the different modes of the fundamental domain encode thermodynamical properties. Partial results on this story are known in the literature \cite{Kruczenski:2005pj}\cite{Bowick:1989us}\cite{Deo:1989bv}\cite{Deo:1988jj}\cite{Spiegelglas:1988hr}, yet the full story and to what extent it can be generalized to curved space has not been studied before. \\
First we look at the strip domain and what its single quantum number $r$ (only winding around one torus cycle) tells us about thermodynamics. This quantum number is the one exhibited in the worldsheet dimensional reduction approach to critical string thermodynamics in section \ref{pathderiv}.\\
We start with the general expression of the partition function of a multiparticle bosonic system:
\begin{equation}
Z = \prod_{k}\frac{1}{1-e^{-\beta E_k}}.
\end{equation}
The free energy of the gas is hence
\begin{equation}
\label{expans}
-\beta F = \sum_{r'=1}^{+\infty}\frac{\sum_k e^{-r'\beta E_k}}{r'}.
\end{equation}
If instead one would assume Maxwell-Boltzmann statistics, we would have $Z = \exp{Z_1}$, which would imply
\begin{equation}
-\beta F = \sum_k e^{-\beta E_k}.
\end{equation}
Comparing the above equations, we see that the $r'=1$ term corresponds to the Maxwell-Boltzmann classical approximation. All higher terms in the Taylor expansion result in Bose-Einstein statistics. 

\subsection{Multistring Instabilities}
Let us look more closely into the structure of the (non-interacting) multistring partition function. For simplicity we only discuss a bosonic system (bosonic string theory) explicitly. The free energy of the string gas can be written as
\begin{equation}
\label{loga}
F= - \frac{1}{\beta}\ln\left(1+Z^{\beta}_1 + Z^{\beta}_2 + \hdots\right),
\end{equation}
where $Z^{\beta}_i$ denotes the partition function for $i$ strings at temperature $\beta^{-1}$. This is also equal to 
\begin{equation}
F= - \frac{1}{\beta}\left(Z^{\beta}_{\left|r\right|=1} + Z^{\beta}_{\left|r\right|=2} + \hdots\right)
\end{equation}
where we consider the single string partition functions for higher winding numbers (strip quantum number $r$). In writing this, we have used the notation $Z^{\beta}_{\left|r\right|=i} = Z^{\beta}_{r=i} + Z^{\beta}_{r=-i}$. The idea is then to series-expand the logarithm in (\ref{loga}) while using relations between the multi-boson partition functions. Let us show this in a bit more detail. Assuming  $Z^{\beta}_1 = Z^{\beta}_{\left|r\right|=1}$ (this is the single-string equivalent of Polchinski's result \cite{Polchinski:1985zf}), the authors of \cite{Kruczenski:2005pj} prove that the next term in the series matches for flat space provided:\footnote{The second order term in the expansion of the logarithm is of the form
\begin{equation}
Z^{\beta}_2 - \frac{1}{2}\left(Z^{\beta}_1\right)^2 = \frac{1}{2}Z^{2\beta}_1,
\end{equation}
where properties of the two-boson partition function were used.}
\begin{equation}
Z^{\beta}_{\left|r\right|=2} = \frac{1}{2}Z^{2\beta}_1,
\end{equation}
which is indeed satisfied for flat space.
As an explicit illustration, we here go one step further and prove that the third term also matches. The partition function for three bosonic strings can in general be written as
\begin{align}
Z^{\beta}_3 &= \sum_{k>l>m}e^{-\beta(E_k+E_l+E_m)} + \sum_{k,l}e^{-\beta(2E_k + E_l)} \\
 &= \frac{1}{6}\left[\sum_{k,l,m}e^{-\beta(E_k+E_l+E_m)} -3\sum_{k,l}e^{-\beta(2E_k+E_l)}+2\sum_ke^{-3\beta E_k} \right] + \sum_{k,l}e^{-\beta(2E_k + E_l)} \\
 &= -\frac{1}{3}\left(Z^{\beta}_{1}\right)^3 + Z^{\beta}_2Z^{\beta}_1 + \frac{1}{3}Z^{3\beta}_1,
\end{align} 
where we used $Z^{2\beta}_1 = 2Z^{\beta}_2 - \left(Z^{\beta}_1\right)^2$. We can rewrite this as
\begin{equation}
\frac{1}{3}Z^{3\beta}_1 = \frac{1}{3}\left(Z^{\beta}_{1}\right)^3 - Z^{\beta}_2Z^{\beta}_1 + Z^{\beta}_3.
\end{equation}
When expanding the logarithm (\ref{loga}), the third order terms are given by
\begin{equation}
Z^{\beta}_3 - Z^{\beta}_2Z^{\beta}_1 + \frac{1}{3}\left(Z^{\beta}_{1}\right)^3.
\end{equation}
The result is hence proven if we can prove that
\begin{equation}
\label{above}
Z^{\beta}_{\left|r\right|=3} = \frac{1}{3}Z^{3\beta}_1.
\end{equation}
This feature is manifest in the explicit flat space free energy expression \cite{Alvarez:1986sj}.\footnote{It is also present in the WZW $AdS_3$ thermal partition function, see for instance equation (31) in \cite{Maldacena:2000kv}. We will discuss more aspects of this model further on.} It also holds in the string path integral derivation sketched in section \ref{pathderiv}.\footnote{When following the derivation in \cite{theory}, the following changes are made. Changing the winding number causes $\beta \to k \beta$ in the string action. The result of the $\tau_1$-integral hence also has the same substitution. However, the integration over the quantum fluctuation $\tilde{X}_0$ gives the same result as before: the periodicity $\beta$ itself is not changed. In all, we get a prefactor $1/\left|k\right|$ to the single string partition function.} Thus in this case we have $Z^{k\beta}_{r=1} = \left|k\right| Z^{\beta}_{r=k}$ for $k\in\mathbb{Z}$ leading to
\begin{equation}
Z^{k\beta}_{\left|r\right|=1} = k Z^{\beta}_{\left|r\right|=k}, \quad k \in \mathbb{N}
\end{equation}
and the above relation (\ref{above}) is a special case of this. The derivation holds now also for curved spaces: the only step that remained to be proven was the above relation and using the string path integral approach of section \ref{pathderiv}, we see that it should hold also for curved spaces. As a first conclusion, it seems that the single quantum number $r$ in the strip should be identified with the $r'$ quantum number in the expansion of the non-interacting bosonic string partition function (\ref{expans}).\footnote{Actually it is the contribution of $+r$ and $-r$ that, upon adding, equals the $r'$ term in the bosonic partition function expansion.} \\

\noindent The benefit of the above discussion is that now we can see explicitly how the multistring partition functions are determined in terms of those containing fewer strings. Take for instance the two-string partition function $Z_2^{\beta}$, which satisfies
\begin{equation}
\frac{1}{2}Z^{2\beta}_{1} = Z^{\beta}_2 - \frac{1}{2}\left(Z^{\beta}_1\right)^2.
\end{equation}
We can read off where this partition function becomes singular: at $T > T_H$, the single particle partition function $Z_1^{\beta}$ diverges. Since $Z^{2\beta}_{1}$ does not diverge, it is clear that the two-string partition function $Z_2^{\beta}$ also diverges. For $T > 2T_H$, if the expression for the partition function can be trusted, an extra singularity develops since now also $Z^{2\beta}_{1}$ diverges. Analogously one can treat the other multiparticle partition sums. What one learns is that at $T=T_H$, not only single-string partition functions diverge, but also all multiparticle sums. In \cite{Spiegelglas:1988hr} it was argued that at $T= lT_H$, the gas becomes unstable for $l$-string states whereas $(l+1)$-string states are still stable. This is interpreted as determining how many long strings are present in the near-Hagedorn gas. Here however we see that multiparticle partition functions all diverge simultaneously at $T=T_H$ whereas formally additional instabilities develop at higher temperatures that are only present in the multistring sector.

\noindent This discussion shows that the number of long strings is not fixed by the temperature (opposed to \cite{Spiegelglas:1988hr}). Thus multi-long-string configurations are also possible and near $T_H$, the gas reconfigures to a system of (possibly multiple) long strings.

\subsection{Interpretation of Hamiltonian quantum numbers in the strip}
\begin{figure}[h]
\centering
\includegraphics[width=12cm]{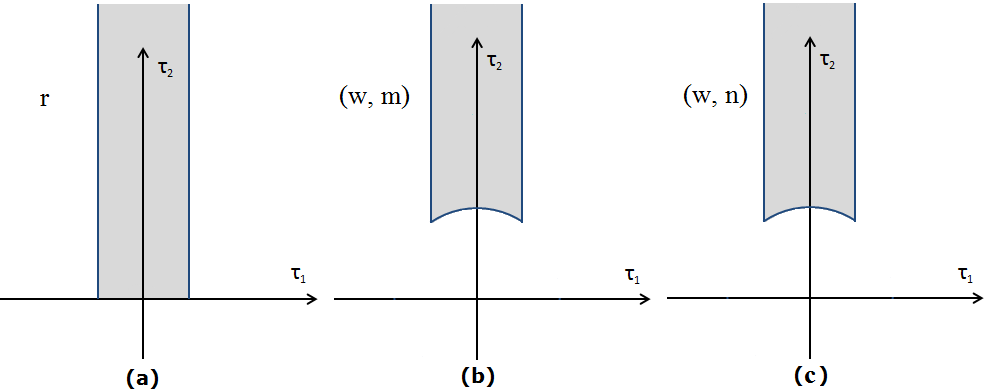}
\caption{(a) The single quantum number $r$ in the strip. (b) The two quantum numbers $w$ and $m$ in the fundamental domain. These appear when performing the torus worldsheet path integral. The link between (a) and (b) is made by utilizing the theorem of \cite{McClain:1986id}\cite{O'Brien:1987pn}. (c) The quantum number $m$ is exchanged for the number $n$ by performing a Poisson resummation. The quantum numbers ($w$, $n$) have meaning in terms of winding and discrete momentum around the compactified time direction and are hence apparent in a Hamiltonian formulation of the theory.}
\label{qn}
\end{figure}

A question that really interests us here is, given a sector ($w$, $n$) in the Hamiltonian (field theory) approach, what does it correspond to in the free energy? In particular, to what part of the strip-sum does it contribute? The answer requires a simple (albeit confusing) number-theoretic reasoning.\\
The translation of the single strip quantum number $r$ and the discrete momentum and winding quantum numbers ($w$, $n$) proceeds in two steps. We shall illustrate these in flat space\footnote{We will come back to curved space further on.} where special attention is paid to analyzing what the contribution is of a specific Hamiltonian sector ($w$, $n$) to any $r$ in the strip. For clarity, the different quantum numbers and their links are shown in figure \ref{qn}.
We will make the link in two steps. In the language of figure \ref{qn}, we first make the link between (a) and (b) and then between (b) and (c).

\subsubsection*{Step 1: Going from (a) to (b)} 
The starting point is the single quantum number $r$ in the strip domain. We start by considering the $\left|r\right|= 1$ contribution. A well-known startegy, is to restrict the modular integral to the fundamental domain while introducing an extra quantum number: we arrive at the double ($w$, $m$) quantum numbers where $m$ and $w$ are coprime \cite{Kruczenski:2005pj} to get the strip $\left|r\right|= 1$ result. These quantum numbers are obtained when one computes the partition function through path integral methods on the fundamental modular domain.
Now we generalize this: we focus on the strip $\left|r\right|= q$ contribution with $q$ an arbitrary positive integer. The coset expansion results in the set ($w$, $m$) with $\frac{m}{q}$ and $\frac{w}{q}$ relatively prime integers. In particular, $m$ and $w$ are multiples of $q$. The different classes exhibited above, can be written as ($w$, $wk + N$) for $N=0\hdots w-1$, for fixed $w$ and $N$, and $k$ runs over $\mathbb{Z}$. In detail, we have for the lowest lying quantum numbers:
\begin{align}
\begin{array}{c|ccccccc}
\hline
F & (w, m) \text{ sectors} \\
\hline
\left|r\right|= 1 & (\pm1,k) & (\pm2, 2k+1)& (\pm3, 3k+1) & (\pm3, 3k+2) & (\pm4, 4k+1) & (\pm4, 4k+3) &\hdots\\
\left|r\right|= 2 & (\pm2, 2k) &  (\pm4, 4k+2) & (\pm6, 6k+2) & (\pm6, 6k+4) &\hdots \\
\left|r\right|= 3 & (\pm3, 3k) & (\pm6, 6k+3) & \hdots \\
\left|r\right|= 4 & (\pm4, 4k) &  \hdots 
\end{array}
\end{align}

\noindent Up to this point, we have merely used the theorem of \cite{McClain:1986id}\cite{O'Brien:1987pn}, but we have paid extra attention to which sectors ($w$, $m$) correspond to a single $r$ quantum number. 

\subsubsection*{Step 2: Going from (b) to (c)}
The second step requires a Poisson resummation on the quantum number $m$. Setting $R_0 = \frac{\beta}{2\pi}$, we remind the reader of the standard textbook result:
\begin{equation}
\label{origi}
\sum_{m\in \mathbb{Z}}\text{exp}\left(-\frac{\pi R_0^2}{\alpha'\tau_2}\left|m-w\tau\right|^2\right) = \sqrt{\frac{\alpha'\tau_2}{R_0^2}}\sum_{n\in\mathbb{Z}}\text{exp}\left(-\pi\tau_2\left(\frac{\alpha' n^2}{R_0^2} + \frac{w^2R_0^2}{\alpha'}\right)+2\pi i \tau_1 nw\right).
\end{equation}
Here we wish to analyze this expression in more detail, tracking the influence of specific subsectors (like we did in the previous subsection). \\
We have discussed above that the transition from one quantum number to two quantum numbers splits these in some classes of the form ($w$, $wk + N$) where $N=0\hdots w-1$. Some of these classes will correspond to the same $r$, but for the moment we do not care about this. For one such sector (fix $w$ and $N$), we have
\begin{equation}
\sum_{m = wk+N, \, k\in\mathbb{Z}}\exp\left(-\frac{\pi R_0^2}{\alpha'\tau_2}\left|m-w\tau\right|^2\right) = \sum_{k=-\infty}^{+\infty}\exp\left(-\frac{\pi R_0^2}{\alpha'\tau_2}\left|(wk+N)-w\tau\right|^2\right),
\end{equation}
which can be Poisson resummed into
\begin{equation}
\label{vg1}
\sqrt{\frac{\alpha'\tau_2}{R_0^2}}\frac{1}{w}\sum_{\tilde{n}=-\infty}^{+\infty}e^{2\pi i \frac{N}{w}\tilde{n}}\exp\left(-\frac{\pi \alpha'\tau_2}{R_0^2w^2}\tilde{n}^2 + 2\pi i \tau_1 \tilde{n} - \frac{\pi R_0^2w^2}{\alpha'\tau_2}\tau_2^2\right).
\end{equation}
An at first sight troubling aspect is the appearance of $\frac{\tilde{n}^2}{w^2}$ in the exponential, suggesting some sort of fractional discrete momentum. Note though that summing the above expression over $N$, restricts $\tilde{n} = w \mathbb{Z}$, such that one obtains (with $\tilde{n} = w n$ and including the sum over $N$) 
\begin{equation}
\sqrt{\frac{\alpha'\tau_2}{R_0^2}}\sum_{n=-\infty}^{+\infty}\exp\left(-\frac{\pi \alpha'\tau_2}{R_0^2}n^2 + 2\pi i \tau_1 n w- \frac{\pi R_0^2w^2}{\alpha'\tau_2}\tau_2^2\right),
\end{equation}
which is indeed the full original result (\ref{origi}). To conclude, the fractional discrete momenta disappear when summing the different sectors and in the end we only care about the non-fractional ones. With this in mind, equation (\ref{vg1}) can be rewritten as
\begin{equation}
\sqrt{\frac{\alpha'\tau_2}{R_0^2}}\frac{1}{w}\sum_{n=-\infty}^{+\infty}\underbrace{e^{2\pi i N n}}_{1}\exp\left(-\frac{\pi \alpha'\tau_2}{R_0^2}n^2 + 2\pi i \tau_1 n w- \frac{\pi R_0^2w^2}{\alpha'\tau_2}\tau_2^2\right).
\end{equation}
What we have found, is that a single sector ($w$, $wk + N$) (for fixed $w$ and $N$) gets contributions from the Hamitonian sectors ($w$, $n$) for all $n$ but with weight $1/w$.

\noindent Turning this conclusion around, the above shows that a Hamiltonian sector ($w$, $n$) contributes equally to the different sectors ($w$, $wk + N$), $N=0\hdots w-1$. For instance, in each sector the $n=0$ mode dominates and contributes equally $\sim 1/w$. Note that all discrete momentum states and winding states correspond to thermal corrections to the free energy and these are \emph{not} the physical particles appearing in the Lorentzian theory. 

\subsubsection*{Final result}
Having found how a Hamiltonian sector ($w$, $n$) contributes to a class of sectors ($w$, $m=wk+N$), we now ask how many of the latter classes of sectors are included in a single $r$ sector.
We have already written down the tabel including the few lowest quantum numbers above. For instance, the $w=4$ states are highlighted below:
\begin{align}
\begin{array}{c|ccccccc}
\hline
F & (w, m) \text{ sectors} \\
\hline
\left|r\right|= 1 & (\pm1,k) & (\pm2, 2k+1)& (\pm3, 3k+1) & (\pm3, 3k+2) & \textcolor{blue}{(\pm4, 4k+1)} & \textcolor{blue}{(\pm4, 4k+3)} &\hdots\\
\left|r\right|= 2 & (\pm2, 2k) &  \textcolor{blue}{(\pm4, 4k+2)} & (\pm6, 6k+2) & (\pm6, 6k+4) &\hdots \\
\left|r\right|= 3 & (\pm3, 3k) & (\pm6, 6k+3) & \hdots \\
\left|r\right|= 4 & \textcolor{blue}{(\pm4, 4k)} &  \hdots 
\end{array}
\end{align}
It is hence clear that a state ($4, n$) for any $n$ contributes with half weight to the $\left|r\right|=1$ sector and with quarter weight to both the $\left|r\right|=2$ and $\left|r\right|=4$ sectors. This procedure can be readily generalized into the following algorithm for finding the contribution of a ($w$, $n$) Hamiltonian state to the different $r$-sectors. 
\begin{itemize}
\item[$(i)$] Find all divisors of $w$.
\item[$(ii)$] For each divisor $i$, find all multiples of $i$ (smaller or equal to $w$) that are \emph{not} multiples of any other larger divisor of $w$ (except 1).
\item[$(iii)$] The number of such multiples for every $i$ is the relative weight to the $\left|r\right|=i$ sector.
\end{itemize}
Note that the same strategy is used for every $n$ and the sum of the weights is $w$. A few examples make this clear. \\
Let us look at the ($w=11$, $n$) state. The above algorithm has the following outcome:
\begin{align}
\begin{array}{|c|c|c|}
\hline
$r$\text{-sector} & \text{multiples} & \text{relative weight} \\
\hline
\left|r\right|=1 & (1,2,3,4,5,6,7,8,9,10) & 10/11 \\
\left|r\right|=11 & (11) & 1/11 \\
\hline
\end{array}
\end{align}
The reason for this simplicity is obviously that 11 is a prime number. A more involved example is the case of a ($w=12$, $n$) state. The above algorithm has the following outcome:
\begin{align}
\begin{array}{|c|c|c|}
\hline
$r$\text{-sector} & \text{multiples} & \text{relative weight} \\
\hline
\left|r\right|=1 & (1,5,7,11) & 4/12 \\
\left|r\right|=2 & (2,10) & 2/12 \\
\left|r\right|=3 & (3,9) & 2/12 \\
\left|r\right|=4 & (4,8) & 2/12 \\
\left|r\right|=6 & (6) & 1/12 \\
\left|r\right|=12 & (12) & 1/12 \\
\hline
\end{array}
\end{align}
Note that in particular, the $w=\pm 1$ state only contributes to the $\left|r\right|=1$ sector with full weight. Hence for these sectors, roughly speaking, the winding number $w$ in the fundamental domain coincides with the number $r$ in the strip, explaining the success we had with equating the strip path integral derivation and the thermal scalar field theory. 

\subsubsection*{Extension to $AdS_3$}
It is interesting to note that the above story applies fully to the $AdS_3$ WZW model as well.\footnote{This model consists of the $AdS_3$ background metric and a Kalb-Ramond background field. The model is an exact string background. An incomplete reference list is the following: \cite{Maldacena:2000kv}\cite{de Boer:1998pp}\cite{Giveon:1998ns}\cite{Kutasov:1999xu}\cite{Gawedzki:1991yu}\cite{Maldacena:2000hw}. Thermodynamics in these spaces has been discussed in \cite{Berkooz:2007fe}\cite{Lin:2007gi}\cite{Mertens:2014nca}.} Firstly, the mapping from strip $r$ to ($w$, $m$) fundamental domain sectors appears generally true: it is explicitly discussed in \cite{Maldacena:2000kv}. Secondly, we have discussed this model in detail in \cite{Mertens:2014nca} and in particular appendix C.2 there allows us to discuss the second step as well. In the notation given there where $l$ takes the role of our $m$, an intermediate step in the computation involves the following Poisson resummation:
\begin{align}
\frac{1}{P}\int_{\mathbb{R}}dm\sum_{l\in\mathbb{Z}}\exp\left(2\pi i \left(\frac{\beta l }{2\pi}m\right)\right) &= \frac{1}{P}\frac{2\pi }{\beta}\int_{\mathbb{R}}dm\sum_{n\in\mathbb{Z}}\delta\left(m-\frac{2\pi n }{\beta}\right).
\end{align}
Focussing now on sectors with $l\to wk + N$ for fixed $N=0\hdots w-1$ and $w$, yields instead for the Poisson resummation step\footnote{The reader is referred to \cite{Mertens:2014nca} for details on where this computation comes from.}
\begin{align}
\frac{1}{P}\int_{\mathbb{R}}dm&\sum_{k\in\mathbb{Z}}\exp\left(2\pi i \frac{\beta m }{2\pi}N\right)\exp\left(2\pi i \left(\frac{\beta kw }{2\pi}m\right)\right) \nonumber \\
&= \frac{1}{P}\frac{2\pi }{w\beta}\int_{\mathbb{R}}dm \exp\left(2\pi i \frac{\beta m }{2\pi}N\right) \sum_{\tilde{n}\in\mathbb{Z}}\delta\left(m-\frac{2\pi \tilde{n} }{w\beta}\right) \nonumber \\
&= \frac{1}{P}\frac{2\pi }{w\beta}\int_{\mathbb{R}}dm \sum_{\tilde{n}\in\mathbb{Z}} \exp\left(2\pi i \frac{\tilde{n} }{w}N\right) \delta\left(m-\frac{2\pi \tilde{n} }{w\beta}\right).
\end{align}
As a check, summing over $N$ gives us a global factor of $w$, and restricts to $\tilde{n} = w \mathbb{Z}$, where we call this new integer $n$ again which will correspond to discrete momentum around the cylinder. \\
Again looking at a fixed sector, we see that if $\tilde{n} \neq w \mathbb{Z}$, the contribution will disappear upon summing. When $\tilde{n} = w \mathbb{Z}$, the phase factor disappears and all sectors contribute equally. The situation is hence exactly as in flat space. It is tempting to speculate that this story actually applies in general to spaces where the thermal circle is topologically stable.

\subsubsection*{Some concluding remarks}
Let us note that the Hamiltonian field theory approach is much better suited to describe subleading corrections to thermodynamical quantities, since these are identified with the next-to-lowest mass state. Such corrections are `dispersed' over the different $r$ sectors from the thermodynamical strip point of view. \\

\noindent The detailed picture exhibited in the two preceding subsections appears more complicated for black hole spacetimes (where the thermal circle pinches off at some point). In \cite{Mertens:2013zya} and \cite{Mertens:2014nca} we discussed that for such spaces not every $w$ mode is present in the thermal spectrum. Therefore the above story cannot hold for such spaces and one needs to look more carefully. Note though that, provided the one-loop thermal path integral corresponds to the free field state counting,\footnote{\label{13}This has been questioned in \cite{Susskind:1994sm}.} the genus one thermal partition function should include all $r$ sectors to satisfy general Bose-Einstein statistics. 

\section{Conclusion}
In a previous paper \cite{theory}, we analyzed the path integral derivation of the random walk behavior of near-Hagedorn thermodynamics \cite{Kruczenski:2005pj}. In this paper, we analyzed several examples that do not have the complications of possible curvature corrections. Examples that do include these have been analyzed elsewhere \cite{Mertens:2013zya}\cite{Mertens:2014nca}. \\
We started by considering the linear dilaton background. This background taught us a valuable lesson concerning the role of the dilaton field in the string path integral and the field theory action: despite treating the dilaton in a totally different way, they agree for on-shell backgrounds as it should be. Also, this background provides an explicit example where continuous quantum numbers in the thermal spectrum are important to determine the critical behavior. The toroidally compactified models showed that also in these spaces (with spatial topological identifications) we reproduce a random walk behavior. The thermal scalar approach also gives us a method to determine the Hagedorn temperature in this space. After that, we considered flat orbifold CFTs to model strings in a box. We learned that the thermal scalar path integral realizes several known topologically non-trivial particle path integrals. String theory provides us with the boundary conditions for the thermal scalar, unlike in particle theories where we are free to choose the boundary conditions. \\
Then we studied a problem of a different nature concerning higher winding modes and their relevance for thermodynamics. We made the link between the Hamiltonian quantum numbers and the single strip quantum number more precise, both for flat space and for an $AdS_3$ model. This suggests these results are actually valid for generic spaces with non-contractible thermal circles. Cigar-shaped thermal manifolds on the other hand do not fall in this category and the resolution of this picture in that case is not known. \\
We believe these examples show that the methods developed in \cite{Kruczenski:2005pj}\cite{theory} lead to suggestive descriptions of the random walk picture directly from the string path integral.

\section*{Acknowledgements}
It is a pleasure to thank David Dudal for several valuable discussions. TM thanks the UGent Special Research Fund for financial support. The work of VIZ was partially supported by the RFBR grant 14-02-01185.

\appendix

\section{Additional computations for the toroidally compactified string}
\subsection{Winding tachyon in string spectrum}
\label{spec}
We show the appearance of a state in the spectrum that becomes massless at a certain temperature in the bosonic string, the superstring and the heterotic string.
\subsubsection*{Bosonic string}
First we show that there is indeed a state in the string spectrum that becomes massless at the Hagedorn temperature. The non-linear sigma model in a general background with constant metric and Kalb-Ramond field has the following mass spectrum \cite{Polchinski:1998rq}
\begin{equation}
m^2 = \frac{1}{2\alpha'^2}G_{mn}\left(v^{m}_Lv^{n}_L+v^{m}_Rv^{n}_R\right) - \frac{4}{\alpha'}
\end{equation}
where $v_{L,R}^{m} = v^{m} \pm w^{m}R_{m}$ and $v_{m} = \frac{\alpha'n_{m}}{R_{m}}-B_{mn}w^{n}R_{n}$. Indices are raised and lowered with the $G_{mn}$ metric.
There is also a constraint
\begin{equation}
n_{m}w^{m}+N-\tilde{N} = 0.
\end{equation}
We are interested in the lowest mass state with non-zero winding in the Euclidean time direction so we set $w^{0} = 1$ and $n_{0}=n_{1}=w^{1}=0.$\footnote{Note that this satisfies the constraint with $N=\tilde{N}=0$.} With the metric and NS-NS field (\ref{metriKR}), we get the following components for the $v$'s:
\begin{align}
v^{0}_{L} = iA(-iB)R_{0}+R_{0}, \quad v^{0}_{R} = iA(-iB)R_{0}-R_{0}, \\
v^{1}_{L} = (1-A^2)(-iB)R_{0}, \quad v^{1}_{R} = (1-A^2)(-iB)R_{0}. 
\end{align}
Inserting this in the mass formula, we get
\begin{equation}
m^2 = \frac{2}{2\alpha'^2}(1-A^2)(1-B^2)\left(R_{0}\right)^2 - \frac{4}{\alpha'}.
\end{equation}
Precisely when 
\begin{equation}
\beta^2 (1-A^2)(1-B^2) = 16\pi^2\alpha'
\end{equation}
we obtain a massless state. This gives us the Hagedorn temperature
\begin{equation}
T_{H} = T_{H,\text{flat}}\sqrt{(1-A^2)(1-B^2)}
\end{equation}
where $T_{H,\text{flat}} = \frac{1}{4\pi\sqrt{\alpha'}}$ is the flat space bosonic Hagedorn temperature. 

\subsubsection*{Type II Superstring}
For type II superstrings exactly the same story holds except for the replacement 
\begin{equation}
\frac{4}{\alpha'} \to \frac{2}{\alpha'}.
\end{equation}
This only modifies the end result in that $T_{H,\text{flat}} \to \frac{1}{2\sqrt{2}\pi\sqrt{\alpha'}}$, the flat space superstring Hagedorn temperature. 

\subsubsection*{Heterotic string}
For heterotic strings, the previous result gets modified in that now the state we are looking at has $n_0 = 1/2$ to satisfy the constraint.
This slightly complicates the derivations:
\begin{align}
v^{0}_{L} = \frac{\alpha'}{2R_{0}} + iA(-iB)R_{0}+R_{0}, \quad v^{0}_{R} = \frac{\alpha'}{2R_{0}} + iA(-iB)R_{0}-R_{0}, \\
v^{1}_{L} = \frac{i\alpha'A}{2R_{0}} + (1-A^2)(-iB)R_{0}, \quad v^{1}_{R} = \frac{i\alpha'A}{2R_{0}} + (1-A^2)(-iB)R_{0}. 
\end{align}
Inserting this in the mass formula (where we now have $\frac{4}{\alpha'} \to \frac{3}{\alpha'}$), we get the following equation for a massless state
\begin{equation}
\beta^2(1-A^2)(1-B^2)+4\pi^2AB\alpha'+\frac{4\pi^4\alpha'^2}{\beta^2}=12\pi^2\alpha'.
\end{equation}
This has two solutions given by
\begin{equation}
\beta_{H\pm}^2 = \frac{6 - 2AB \pm 2\sqrt{8-6AB+A^2+B^2}}{(1-A^2)(1-B^2)}\pi^2\alpha'.
\end{equation}
Just like in the uncompactified case, the physical solution is the one with the plus sign. \\

\subsection{Path integral and summation-of-particles are equivalent for the free energy}
\label{sop}
We now only consider the bosonic string. Using the summation-of-particles strategy to calculate the free energy, the authors of \cite{Grignani:2001ik} found that the free energy is given by
\begin{align}
F = &- \sum_{n,p,l}\int_{0}^{+\infty}\frac{d\tau_2}{2\tau_2}\int_{-1/2}^{1/2}\frac{d\tau_1}{(4\pi^2\alpha'\tau_2)^{13}}\left(\frac{\alpha'\tau_2}{R_{1}^2}\right)^{1/2}\left|\eta(\tau)\right|^{-48} \nonumber\\
&\times\exp\left[-\frac{\beta^2n^2}{4\pi\alpha'\tau_2}-\pi\alpha'\tau_2\left(\frac{l^2}{R_{1}^2}+\frac{R_{1}^2p^2}{\alpha'^2}\right)-2\pi i \tau_1pl+n\beta B\frac{R_{1}p}{\alpha'} + n\beta A\frac{l}{R_{1}}\right].
\end{align}

\noindent We show that this can also be found as a torus path integral where we choose the strip as domain and we restrict attention to torus winding for $X^0$ only in the $\sigma_2$ direction. The $X^{1}$ target field includes winding in both torus cycles. So this is an explicit check that the free energy is equal to a path integral on the thermal manifold for this set-up. \\
We return to the original genus one string action (before any modular transformation)
\begin{eqnarray}
S = \frac{1}{4\pi\alpha'}\int_0^1d\sigma_1\int_0^1d\sigma_2\left[\frac{\tau_1^2+\tau_2^2}{\tau_2}\partial_1X^{\mu}\partial_1X^{\nu}G_{\mu\nu}-2\frac{\tau_1}{\tau_2}\partial_1X^{\mu}\partial_2X^{\nu}G_{\mu\nu} \right. \nonumber\\
\left. +\frac{1}{\tau_2}\partial_2X^{\mu}\partial_2X^{\nu}G_{\mu\nu}+2i\partial_1X^{\mu}\partial_2X^{\nu}B_{\mu\nu}\right].
\end{eqnarray}
with the expansions
\begin{align}
X^0(\sigma,\tau) & =  n \beta \sigma_2 +  \text{periodic}, \nonumber \\
X^1(\sigma,\tau) & =  p\beta_1\sigma_1 + m\beta_1\sigma_2 + \text{periodic}, \nonumber \\
X^i(\sigma,\tau) & =  \text{periodic}. 
\end{align}
Inserting this in the torus path integral, we get the non-oscillator contribution
\begin{eqnarray}
S_{non-osc} = \frac{1}{4\pi\alpha'}\left[\frac{\tau_1^2+\tau_2^2}{\tau_2}p^2\beta_1^2 - 2\frac{\tau_1}{\tau_2}pm\beta_1^2 + 2i\frac{\tau_1}{\tau_2}\beta\beta_1Anp + \frac{1}{\tau_2}m^2\beta_1^2 \right. \nonumber \\
\left.  + \frac{1}{\tau_2}(1-A^2)n^2\beta^2 - 2i\frac{1}{\tau_2}\beta\beta_1Anm-2B\beta\beta_1 np\right].
\end{eqnarray}
Using a Poisson resummation in $m$, we obtain precisely the previous expression (after some straightforward arithmetics and including the oscillator path integral).

\subsection{Most general flat toroidal model}
\label{generalization}
The results from section \ref{torcompmodel} can be readily generalized to the most general flat space toroidal model with an arbitrary constant metric $G_{mn}$ and arbitrary constant Kalb-Ramond field $B_{mn}$ with $m,n=0\hdots D-1$. 
Firstly, the state in the thermal spectrum is readily found as we now discuss. The general mass formula is given by \cite{Polchinski:1998rq}:
\begin{equation}
m^2 = \frac{1}{2\alpha'^2}G_{mn}\left(v^{m}_Lv^{n}_L+v^{m}_Rv^{n}_R\right) - \frac{4}{\alpha'},
\end{equation}
where $v_{L,R}^{m} = v^{m} \pm w^{m}R_{m}$ and $v_{m} = \frac{\alpha'n_{m}}{R_{m}}-B_{mn}w^{n}R_{n}$. 
For a general toroidal background, we have
\begin{align}
v_0 = 0, &\quad v_i = -B_{i0}R_0, \\
v^{0} = -G^{0i}B_{i0}R_0, &\quad v^{i} = -G^{ij}B_{j0}R_0, \\
v^{0}_{L,R} = -G^{0i}B_{i0}R_0 \pm R_0, &\quad v^{i}_{L,R} = -G^{ij}B_{j0}R_0,
\end{align}
and so
\begin{equation}
m^2 = \frac{\beta^2}{4\pi^2\alpha'^2}\left[G_{00}G^{0i}G^{0j}B_{i0}B_{j0} + G_{00} + 2G_{0i}G^{0j}G^{ik}B_{j0}B_{k0} + G_{ij}G^{il}G^{jk}B_{l0}B_{k0}\right]- \frac{4}{\alpha'},
\end{equation}
which can be simplified into
\begin{equation}
m^2 = \frac{\beta^2}{4\pi^2\alpha'^2}\left[ G_{00} + G^{lk}B_{l0}B_{k0}\right]- \frac{4}{\alpha'}.
\end{equation}
In the path integral language, the same steps as in section \ref{torcompmodel} are used. We briefly sketch the intermediate results.
\begin{itemize}
\item In the $\tau_2 \to \infty$ limit, all winding contributions around the $\sigma$-direction are dominated by $w_{i}=0$. This generalizes the case in section \ref{torcompmodel} where $w=0$ was shown to dominate.
\item The prefactors combine into $\prod_{i} \left(\frac{\alpha'\tau_2}{R_i^2}\right)^{1/2}$. After the coordinate redefinition, the flat space coordinates live in the toroidal region of size $\prod_{i}R_{i}$.
\item One needs to do $D$ Poisson resummations and then the $\tau_1$ integration. This results in 
\begin{equation}
S = \frac{\tau_2 \beta^2}{4\pi\alpha'}\left[G_{00} + \sum_{i,j}\frac{B_{i0}B_{j0}M_{0i}M_{0j}}{(\det G) M_{00}} + \sum_{i,j}\frac{B_{i0}B_{j0}M_{0i,j0}}{M_{00}}\right]
\end{equation}
where $M_{ab}$ denotes the cofactor of the matrix $G_{\mu\nu}$ and $M_{ab,cd}$ denotes the minor where rows $a$ and $c$ and columns $b$ and $d$ are deleted (with the sign $(-1)^{a+b+c+d}$).
In general the following identity is valid:\footnote{This formula is slightly more general than the so-called Desnanot-Jacobi identity, but is still a special case of the Jacobi identity \cite{Gradshteyn:2007}. }  
\begin{equation}
\frac{M_{0i}M_{0j}}{(\det G)M_{00}} + \frac{M_{0i,j0}}{M_{00}} = \frac{M_{ij}}{\det G}.
\end{equation}
This results in 
\begin{align}
S &= \frac{\tau_2 \beta^2}{4\pi\alpha'}\left[G_{00} + \sum_{i,j}\frac{M_{ij}B_{i0}B_{j0}}{\det G}\right] \\
 &=\frac{\tau_2 \beta^2}{4\pi\alpha'}\left[G_{00} + \sum_{i,j}G^{ij}B_{i0}B_{j0}\right]. 
\end{align}
This action leads to the same Hagedorn temperature as the previous calculation
\begin{equation}
T_{H} = T_{H,\text{flat}}\sqrt{G_{00} + G^{lk}B_{l0}B_{k0}}.
\end{equation}
\end{itemize}
We make several remarks concerning this result.
\begin{itemize}
\item 
Note that the mixed components $G_{0i}$ and $B_{0i}$ are both imaginary while $G_{00}$ is real and positive. The above Hagedorn temperature makes sense since $\det G$ and hence $G^{lk}$ is real so $G_{00} + G^{lk}B_{l0}B_{k0}$ is indeed a real number. It might become negative, but this simply indicates the presence of critical backgrounds, as discussed in \cite{Grignani:2001ik} and in section \ref{interpre}.
\item 
If $B_{\mu\nu} = 0$, only the metric component $G_{00}$ contributes. In particular, as long as the $i0$ components are zero, the Hagedorn temperature is unaffected (except the $G_{00}$ component). This result is more general: it holds also for smooth compactifications (with unitary CFTs) as long as the space factorizes. Note also that the Hagedorn temperature is independent of $B_{ij}$.
\item 
For type II superstrings, one again simply substitutes the correct value of $T_{H,\text{flat}}$ to find the Hagedorn temperature. 
\item
For heterotic strings, two extra terms appear in the action
\begin{align}
S_1 &= -\pi \tau_2 \sum_i \frac{B_{i0}M_{0i}}{\det G} = -\pi \tau_2 \sum_i B_{i0}G^{0i}, \\
S_2 &= \frac{\pi^3\alpha'\tau_2}{\beta^2}\frac{M_{00}}{\det G} =\frac{\pi^3\alpha'\tau_2}{\beta^2}G^{00}.
\end{align}
Explicitly working out the mass$^2$ in the thermal spectrum also gives precisely these two terms.
Finally one readily finds the Hagedorn temperature
\begin{equation}
\mbox{{\small{$\displaystyle\frac{\beta_H^2}{2\pi^2\alpha'} = \frac{3 + \sum_i B_{i0}G^{0i}+\sqrt{9 + 6\sum_i B_{i0}G^{0i} + \sum_{i,j}B_{i0}B_{j0}\left(G^{i0}G^{j0}-G^{ij}G^{00}\right)-G_{00}G^{00}}}{G_{00}+\sum_{i,j}G^{ij}B_{i0}B_{j0}}$}}}.
\end{equation} 
\end{itemize}
These results demonstrate again that a divergence in the partition function (path integral calculation) matches directly to a marginal state in the Euclidean spectrum.

\end{document}